\pdfoutput=1

\documentclass[11pt]{article}

\usepackage[final]{acl}

\usepackage{times}
\usepackage{latexsym}
\usepackage{amsmath}
\usepackage{subfigure}
\usepackage{algorithm, algorithmic}
\usepackage{amsfonts}
\usepackage{amssymb}
\usepackage{booktabs}
\usepackage{multirow}
\usepackage{colortbl}
\usepackage{xcolor}
\usepackage{enumitem}

\usepackage[T1]{fontenc}

\usepackage[utf8]{inputenc}

\usepackage{microtype}

\usepackage{inconsolata}

\usepackage{graphicx}

\title{Taylor Unswift: Secured Weight Release for \\Large Language Models via Taylor Expansion}

\author{
 \textbf{Guanchu Wang\textsuperscript{1}}\thanks{Equal contribution, ordered by rolling dices.},
 \textbf{Yu-Neng Chuang\textsuperscript{1}$^*$},
 \textbf{Ruixiang Tang\textsuperscript{2}},
 \textbf{Shaochen Zhong\textsuperscript{1}},
 \textbf{Jiayi Yuan\textsuperscript{1}},
 \\
 \textbf{Hongye Jin\textsuperscript{3}},
 \textbf{Zirui Liu\textsuperscript{5}},
 \textbf{Vipin Chaudhary\textsuperscript{4}},
 \textbf{Shuai Xu\textsuperscript{4}},
 \textbf{James Caverlee\textsuperscript{3}},
 \textbf{Xia Hu\textsuperscript{1}}
\\
\\
 \textsuperscript{1}Rice University,
 \textsuperscript{2}Rutgers University,
 \textsuperscript{3}Texas A\&M University,
\\
 \textsuperscript{4}Case Western Reserve University
  \textsuperscript{5}University of Minnesota
\\
 \small{
 \texttt{\{gw22,yc146,henry.zhong,jy101,xia.hu\}@rice.edu; ruixiang.tang@rutgers.edu;}}
 \\
 \small{
 \texttt{\{jhy0410,caverlee\}@tamu.edu;}
 \texttt{\{vipin,sxx214\}@case.edu;}
 }
  \small{
 \texttt{zrliu@umn.edu;}
 }
}

\def\Algname{Taylor-series MLP}
\def\Algnameunderline{\textcolor{darkred}{Taylor}-series \textcolor{darkred}{MLP}}
\def\Algnameabbr{TaylorMLP}

\definecolor{lightred}{RGB}{247, 220, 220}
\definecolor{lightgreen}{RGB}{220, 247, 227}
\definecolor{darkred}{HTML}{be002f}
\definecolor{darkgreen}{HTML}{00a64f}

\begin{document}
\maketitle
\begin{abstract}


Ensuring the security of released large language models (LLMs) poses a significant dilemma, as existing mechanisms either compromise ownership rights or raise data privacy concerns. To address this dilemma, we introduce \Algnameabbr{} to protect the ownership of released LLMs and prevent their abuse. Specifically, \Algnameabbr{} preserves the ownership of LLMs by transforming the weights of LLMs into parameters of Taylor-series. Instead of releasing the original weights, developers can release the Taylor-series parameters with users, thereby ensuring the security of LLMs. Moreover, \Algnameabbr{} can prevent abuse of LLMs by adjusting the generation speed. It can induce low-speed token generation for the protected LLMs by increasing the terms in the Taylor-series. This intentional delay helps LLM developers prevent potential large-scale unauthorized uses of their models. Empirical experiments across five datasets and three LLM architectures demonstrate that \Algnameabbr{} induces over $\mathbf{4 \times}$ increase in latency, producing the tokens precisely matched with original LLMs. Subsequent defensive experiments further confirm that \Algnameabbr{} effectively prevents users from reconstructing the weight values based on downstream datasets.
The source code is available at \url{https://github.com/guanchuwang/Taylor-Unswift}.

\end{abstract}

\section{Introduction}


Training large language models (LLMs) is an expensive and complex endeavor, requiring substantial investments in financial and computational resources~\cite{wei2021finetuned}. 
In particular, it requires a huge collection of high-quality datasets from diverse domains, which proves to be labor-intensive and time-consuming\nocite{longpre2023flan, cui2024scgpt, yuan2023llm, wu2023pmc}.
However, once released, the LLMs may face the risks of abuse such as unethical or commercial exploitation~\cite{weidinger2021ethical}.
This raises the critical and urgent challenge of ensuring the security of released LLMs.



\begin{figure}[t]
    \centering
    \subfigcapskip=-2mm
    \!\!\!\!\!\!\!\!  \subfigure[API released LLMs cause data privacy concerns.]{\includegraphics[width=0.99\linewidth]{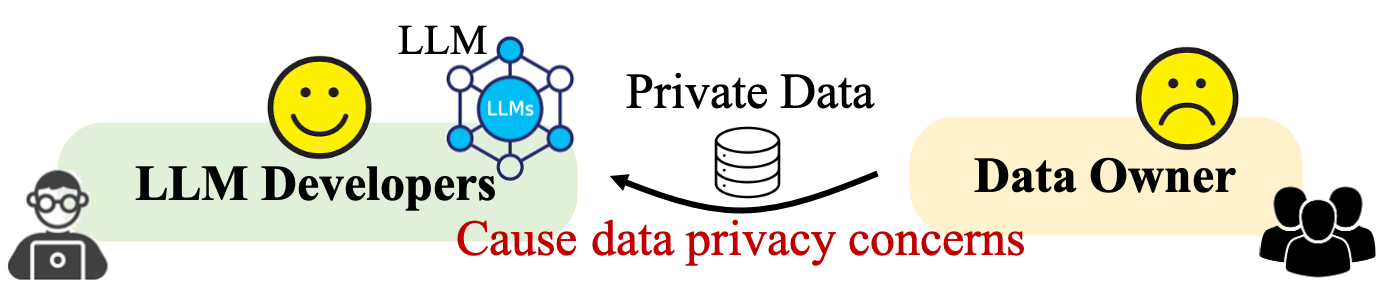}}
    \vspace{-2mm}
    \\
    \!\!\!\!\!\!\!\!  \subfigure[Open-source breaks the ownership of LLMs.]{\includegraphics[width=0.99\linewidth]{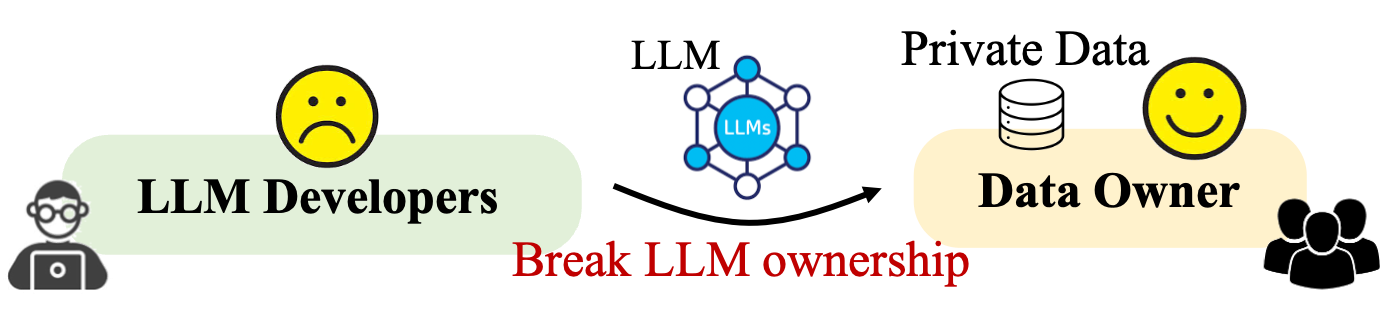}}
    \vspace{-2mm}
    \\
    \subfigure[\Algnameabbr{} protects LLM ownership and data privacy.]{\includegraphics[width=0.99\linewidth]{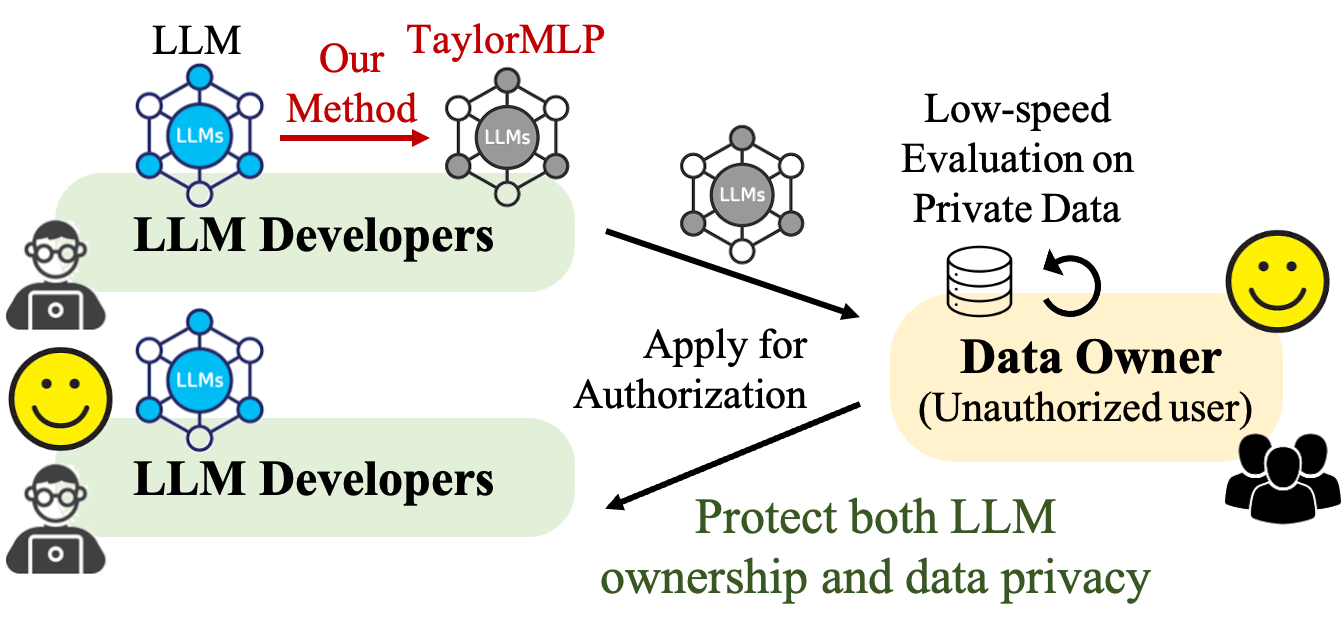}}
    \vspace{-4mm}
    \caption{\small \label{fig:LLM_access} Existing mechanisms for releasing LLMs: (a) API release and (b) open-source. We propose (c) \Algnameabbr{} to protect the ownership of released LLMs.}
    \vspace{-4mm}
\end{figure}




In this work, we explore the security aspects of existing mechanisms for releasing LLMs. This process involves developers providing access to their models for general users, with criteria restricting access to ethical and non-commercial purposes~\cite{wang2024assessing}. 
Currently, there are two primary mechanisms for releasing LLMs: \textit{\textbf{API Release}}~\cite{caruccio2024claude} and \textit{\textbf{Open-source}}~\cite{raffel2020exploring}.
With the release of APIs, user authorization is managed through specific API keys. 
Examples of this mechanism include the ChatGPT\cite{achiam2023gpt}, Gemini~\cite{team2023gemini}, and Claude models~\cite{caruccio2024claude}.
In these cases, users do not have access to the architectures or weights of the LLMs. Instead, they share their private data to the developers and receive the processed results, as shown in Figure~\ref{fig:LLM_access}~(a). Consequently, \textit{\textbf{the API release mechanism may cause the data privacy concerns}}~\cite{yang2023harnessing}.
On the other hand, the open-source mechanism fully shares the LLM weights with users, as shown in Figure~\ref{fig:LLM_access}~(b)~\cite{wolf2019huggingface}. 
Common examples include Llama~\cite{touvron2023llama}, Mixtral~\cite{jiang2024mixtral}, and Phi~\cite{team2024gemma}. 
While the open-source mechanism ensures the safety of users' private data, it also raises significant challenges for developers. 
Specifically, \textit{\textbf{the open-source mechanism can break the ownership of LLMs}}, as users gain control over the models and can use them for any purpose, even those prohibited by the developers. 
For example, users may exploit LLMs for unethical or commercial purposes, both of which may be prohibited by the developers.
This potential loss of control and ownership may lead many model developers to avoid sharing their models ~\cite{zha2023data, sharir2020cost}. Therefore, \textbf{\textit{there is a dilemma between protecting ownership rights and ensuring open access to LLMs}}, posing a significant challenge for developers.


\subsection*{Can we solve the access dilemma for LLMs?}

We propose \Algnameunderline{}~(\Algnameabbr{}) to protect the ownership of released LLMs and prevent their potential abuse.
As illustrated in Figure~\ref{fig:LLM_access}~(c), \Algnameabbr{} addresses the dilemma by securing the weights of LLMs into latent parameters. By sharing these parameters instead of the original weights with users, developers can maintain ownership of their models while allowing users to harness the model's performance. Specifically, \Algnameabbr{} converts the original weights into parameters of the Taylor series. Our empirical experiments confirm that it is infeasible to reconstruct the original weights from these Taylor series parameters, thereby ensuring the security of the model's parameters and allowing safe access to its functional capabilities without full model exposure.





\subsection*{Can we prevent unauthorized users from exploiting the LLM for their own purposes?}

To prevent unauthorized users from abusing the protected LLMs, \Algnameabbr{} allows developers to control the utility of the LLM by adjusting the speed of token generation.
Specifically, \Algnameabbr{} induces \textit{\textbf{low-speed token generation}} for the secured LLMs by increasing the terms in the Taylor-series.
It significantly increases the number of floating-point operations required for the generation process, leading to a notable increase in latency.
Our empirical studies show that \Algnameabbr{} induces more than $\mathbf{4 \times}$ increases in latency, while maintaining the produced tokens precisely matched with original LLMs. This intentional delay helps developers prevent the potential large-scale unauthorized use of their released LLMs.

\subsection*{How does \Algnameabbr{} perform in practice?}

To evaluate \Algnameabbr{}, we conducted experiments across five datasets: TruthfulQA, MathQA, MMLU, OpenbookQA, and Wikitext-2; and three different LLM architectures: Llama-3-8B, Mistral-7B, and Phi-2. 
The experimental results demonstrate that \Algnameabbr{} fully retains the accuracy and chat capabilities of original LLMs, while inducing $\mathbf{4 \times} \! \sim \! \mathbf{8 \times}$ increases in latency of token generation.
Subsequent defensive experiments further confirm that \Algnameabbr{} effectively prevents users from reconstructing the weight values based on downstream datasets.
In summary, our work makes the following contributions:
\begin{itemize} [leftmargin=4mm, topsep=2mm]

    \item \textbf{Preserving LLM Ownership.} 
    \Algnameabbr{} preserves the ownership of LLM by transforming the weights of LLMs into parameters of Taylor-series. It is infeasible to reconstruct the weights from the Taylor-series parameters. 
    
    
    

    
    \item \textbf{Preventing Abuse.}
    \Algnameabbr{} prevents abuse of LLMs by adjusting the generation speed. It induces low-speed token generation process, preventing the potential large-scale unauthorized use of the protected LLMs.
    
    \item \textbf{Evaluation.} Experiment results across five datasets and three LLM architectures show that \Algnameabbr{} retains the accuracy and chat capabilities while preventing reconstruction of the weight values based on downstream datasets.



    
\end{itemize}

\section{Architecture of Transformers}

\begin{figure}
    \centering
    \subfigcapskip=-2mm
    \subfigure[Original MLP layers]{
    \includegraphics[width=0.9\linewidth]{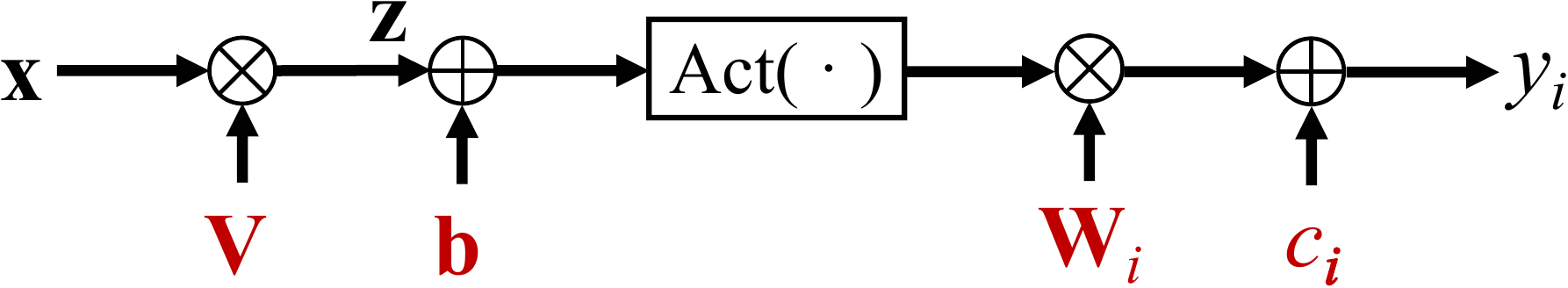}
    }
    \subfigure[\Algnameabbr{} layers]{
    \includegraphics[width=0.9\linewidth]{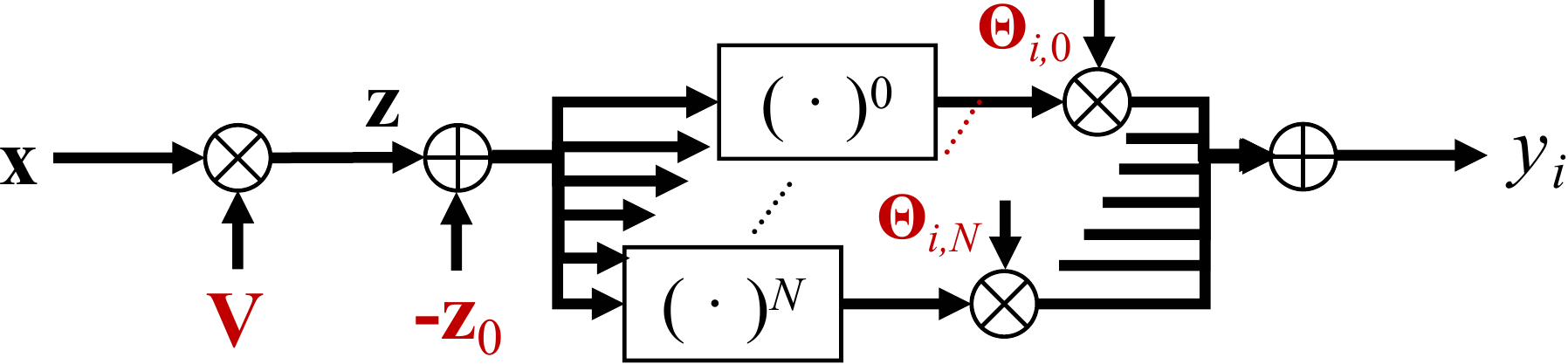}
    }
    \vspace{-2mm}
    \caption{(a) Original MLP layers parameterized by \textcolor{darkred}{$\textbf{V}, \textbf{b}, \textbf{W}$}, and \textcolor{darkred}{$\textbf{c}$}. (b) \Algnameabbr{} layers. \Algnameabbr{} converts $\textbf{b}, \textbf{W}_i$ and $c_i$ into \textcolor{darkred}{$\{\boldsymbol{\Theta}_{i,0}, \cdots, \boldsymbol{\Theta}_{i,N}\}$} for securing their values. This process is irreversible.}
    \label{fig:protect_mlp}
    \vspace{-4mm}
\end{figure}

We focus on Transformer-based LLMs~\cite{vaswani2017attention, brown2020language} to develop the method of LLM protection.
A transformer block consists of an attention layer and MLP layer.
An MLP layer is a pipeline of a linear layer, activation function, and another linear layer, whose architecture is shown in Figure~\ref{fig:protect_mlp}~(a).
Let $\textbf{V}, \textbf{b}$ and $\textbf{W}, \textbf{c}$ denote the weight matrixes of the two linear layers.
Given the input tensor $\textbf{x}$ to the MLP layer, the output value in the $i$-th dimension is given by
\begin{align}
\label{eq:mlp_inference}
y_i = \big\langle \textbf{W}_i, \text{Act}(\textbf{z} + \textbf{b}) \big\rangle + c_i,
\end{align}
where $\textbf{z} = \textbf{V}\textbf{x}$; $\langle \bullet, \bullet \rangle$ denotes the inner product; and $\textbf{W}_i$ and $c_i$ denote the $i$-th row and $i$-th element of $\textbf{W}$ and $\textbf{c}$, respectively.
In this work, we introduce a method to secure the weights of MLP layers within the transformer architecture.
This can be widely applied to powerful LLMs, such as Llama~\cite{touvron2023llama}, Phi~\cite{gunasekar2023textbooks}, and Gemma~\cite{team2024gemma}.


\section{\Algname{}~(\Algnameabbr{})}


The framework of \Algnameabbr{} is shown in Figure~\ref{fig:protect_mlp}~(b).
Intuitively, \Algnameabbr{} transforms the weight matrices $\textbf{b}, \textbf{W}$ and $\textbf{c}$ into latent parameters ${\boldsymbol{\Theta}_{i,0}, \! \cdots \!, \! \boldsymbol{\Theta}_{i,N}}$.
In this way, \Algnameabbr{} generates output tokens without needing the original weight values.
In this section, we first describe the weight transformation of \Algnameabbr{}.
Then, we demonstrate the outputs of \Algnameabbr{} theoretically converges to the original MLPs.
Finally, we discuss the benefits of \Algnameabbr{} to LLM community. 


\subsection{Securing MLP Weights by Taylor-series}

\Algnameabbr{} transforms the weight matrix $\textbf{b}, \textbf{W}$ and $\textbf{c}$ into the latent space based on the Taylor Expansion Theory.
Specifically, based on a local embedding $\textbf{z}_0$, the $\text{Act}(\textbf{z} + \textbf{b})$ term in Equation~(\ref{eq:mlp_inference}) can be reformulated into the Taylor-series as follows:
\begin{align}
\label{eq:act_taylor}
\text{Act}(\textbf{z} \!+\! \textbf{b}) \!\approx\! \sum_{n=0}^{N} \frac{\text{Act}^{(n)}(\textbf{z}_0 \!+\! \textbf{b})}{n !} \!\odot\! (\textbf{z} \!-\! \textbf{z}_0)^n, \!\!\!\!\!\!
\end{align}
where $(\textbf{z} \!-\! \textbf{z}_0)^n$ indicates the element-wise power of $n$; $\text{Act}^{(n)}(\bullet)$ denotes the $n$-order derivative of $\text{Act}(\bullet)$.
Representative activation functions for the Transformers are $\text{GELU}(\bullet)$~\cite{vaswani2017attention} and $\text{SiLU}(\bullet)$~\cite{touvron2023llama}, whose $n$-order derivative are given in Appendixes~\ref{appendix:gelu_grad} and~\ref{appendix:silu_grad}.


Following the Taylor-series in Equation~(\ref{eq:act_taylor}), the forward pass of MLP layers can be reformulated to eliminate the original weight matrices.
Specifically, by applying the Taylor expansion of $\text{Act}(\textbf{z} + \textbf{b})$ from Equation~(\ref{eq:act_taylor}) to Equation~(\ref{eq:mlp_inference}), the forward pass can be reformulated to:
\begin{align}
y_i &\approx \Big\langle \textbf{W}_i, \sum_{n=0}^{N} \frac{\text{Act}^{(n)}(\textbf{z}_0 + \textbf{b})}{n !} \odot (\textbf{z} - \textbf{z}_0)^n \Big\rangle + c_i
\nonumber
\\
&= \! \sum_{n=0}^{N} \! \Big\langle \textbf{W}_i \!\odot\! \text{Act}^{(n)}(\textbf{z}_0 + \textbf{b})(n !)^{-1}\!\!, (\textbf{z} - \textbf{z}_0)^n \Big\rangle,
\nonumber
\\
\label{eq:talor_forward}
&\triangleq \sum_{n=0}^{N} \Big\langle \boldsymbol{\Theta}_{i,n}, (\textbf{z} - \textbf{z}_0)^n \Big\rangle,
\end{align}
where $\{\boldsymbol{\Theta}_{i,0}, \cdots, \boldsymbol{\Theta}_{i,N}\}$ denote the parameters in the latent space; and $\boldsymbol{\Theta}_{i,n}$ depends on the $n$-order derivative $\text{Act}^{(n)}(\textbf{z}_0 + \textbf{b})$, as given by
\begin{align}
\boldsymbol{\Theta}_{i,0} &= \textbf{W}_i \odot \text{Act}(\textbf{z}_0 + \textbf{b}) + c_i
\nonumber
\\
\label{eq:MSP_weight}
\boldsymbol{\Theta}_{i,n} &= \textbf{W}_i \odot \text{Act}^{(n)}(\textbf{z}_0 + \textbf{b})(n !)^{-1}.
\end{align}
After the transformation, \Algnameabbr{} eliminates the need for $\textbf{b}$, $\textbf{W}$, and $\textbf{c}$ in the generation process. 
Moreover, it cannot reconstruct $\textbf{b}$, $\textbf{W}$, and $\textbf{c}$ from ${\boldsymbol{\Theta}_{i,0}, \cdots, \boldsymbol{\Theta}_{i,N}}$, thereby securing the original weight matrices.
\Algnameabbr{} can be applied to protect the MLP layers within LLMs.



\subsection{Estimating the Local Embedding $\textbf{z}_0$}

We clarify the local embedding $\textbf{z}_0$.
Specifically, to minimize the difference between the outputs of \Algnameabbr{} and original MLPs, we minimize the difference between the two sides of Equation~(\ref{eq:act_taylor}). 
This is equivalent to minimizing the distance between $\textbf{z}_0$ and the embedding $\textbf{z}$, where $\textbf{z} = \textbf{V}\textbf{x}$ for the testing input $\textbf{x}$.
We address this problem by taking $\textbf{x}$ from large-scale datasets $\mathcal{D}$ and applying the following solution: 
\begin{align}
\label{eq:local_embedding}
\textbf{z}_0^* &= \arg\min \max_{\textbf{x} \sim \mathcal{D}} || \textbf{z} - \textbf{z}_0 ||_1
\\
&= \frac{1}{2} (\textbf{z}_{\text{max}} + \textbf{z}_{\text{min}}),
\end{align}
where the $i$-th element of $\textbf{z}_{\text{max}}$ takes the value of $\max_{\textbf{x} \sim \mathcal{D}} \textbf{x}^{\mathsf{T}}\textbf{V}[:, i]$; that of $\textbf{z}_{\text{min}}$ takes $\min_{\textbf{x} \sim \mathcal{D}} \textbf{x}^{\mathsf{T}}\textbf{V}[:, i]$.
The optimal $\textbf{z}_0^*$ can be generally effective for downstream tasks if $\mathcal{D}$ is sufficiently large.
In our work, we take the large-scale pile datasets\footnote{\url{https://huggingface.co/datasets/mit-han-lab/pile-val-backup}} as $\mathcal{D}$ for the estimation of $\textbf{z}_0$.


\subsection{Algorithm}

The algorithm of \Algnameabbr{} is given in Algorithm~\ref{alg:protect}.
Specifically, given the weights $\textbf{b}, \textbf{W}$, $\textbf{c}$, and local embedding $\textbf{z}_0$, \Algnameabbr{} follows Equation~(\ref{eq:MSP_weight}) to transform them into the latent values~(lines~3 and 6).
After the transformation, \Algnameabbr{} can generate output tokens by $y_i = \sum_{n=0}^{N} \big\langle \boldsymbol{\Theta}_{i,n}, (\textbf{z} - \textbf{z}_0)^n \big\rangle$, as shown in Figure~\ref{fig:protect_mlp}~(b).

\begin{algorithm}[t] %
\caption{Transforming MLP to \Algnameabbr{}.}
\label{alg:protect}
\textbf{Input:} $\text{MLP}(\bullet ~|~ \textbf{V}, \textbf{b}, \textbf{W}, \textbf{c})$ and $\textbf{z}_0$  \\
\textbf{Output:} {$\text{\Algnameabbr{}} (\bullet | \textbf{V}, \! \textbf{z}_0, \! \{\boldsymbol{\Theta}_{i,0}, \! \cdots, \! \boldsymbol{\Theta}_{i,N}\}_{i=1}^{D} )$} \\
\vspace{-5mm}
\begin{algorithmic}[1]

\FOR{$i := 1 \text{ to } D$}  

\STATE $\textbf{W}_i$ and $c_i$ take the $i$-th row and $i$-th element of $\textbf{W}$ and $\textbf{c}$, respectively.
\STATE $\boldsymbol{\Theta}_{i,0} = \textbf{W}_i \odot \text{Act}(\textbf{z}_0 + \textbf{b}) + c_i$

\FOR{$n := 1 \text{ to } N$}  

\STATE $\boldsymbol{\Theta}_{i,n} = \textbf{W}_i \odot \text{Act}^{(n)}(\textbf{z}_0 + \textbf{b})(n !)^{-1}$

\ENDFOR
\ENDFOR
\end{algorithmic}
\end{algorithm}





\subsection{Theoretical Convergence of \Algnameabbr{}}
\label{sec:theory}

In this section, we theoretically demonstrate the output of \Algnameabbr{} converges to that of original MLP when $N \to \infty$.
For $\forall \textbf{z} \in \mathbb{R}^{D}$, the sum of Taylor series theoretically converge to the output of activation function, which is given by
\begin{align}
\lim_{N \to \infty} \sum_{n=0}^{N} \frac{\text{Act}^{(n)}(\textbf{z}_0 \!+\! \textbf{b})}{n !} \odot (\textbf{z} \!-\! \textbf{z}_0)^n \!=\! \text{Act}(\textbf{z} \!+\! \textbf{b})
\nonumber
\end{align}
This enables the value of Equation~(\ref{eq:talor_forward}) to converge to Equation~(\ref{eq:mlp_inference}).
Consequently, we have the output of \Algnameabbr{} converge to that of original MLPs when $N \to \infty$, given as follows:
\begin{align}
&\lim_{N\to \infty} \! \text{\Algnameabbr{}} \big(\bullet \! ~|~ \textbf{V}, \textbf{z}_0, \{\boldsymbol{\Theta}_{i,j}\}_{1\leq i\leq D, 0\leq j\leq N} \big)
\nonumber
\\
&= \text{MLP}(\bullet ~|~ \textbf{V}, \textbf{b}, \textbf{W}, \textbf{c})
\nonumber
\end{align}

Note that we cannot have ${N\to \infty}$ in practice.
We show in Section~\ref{sec:exp_expan_order} that $N \geq 8$ is sufficiently large for converging to original MLP outputs.







\begin{table*}[]
    \centering
\caption{Accuracy and per-token latency of \Algnameabbr{} compared with the results of original LLMs. Numbers in \textcolor{red}{red} and \textcolor{darkgreen}{green} indicate failures and successes of retaining the original accuracy, respectively.}
\vspace{-3mm}
    \label{tab:acc_latency}
    \resizebox{\textwidth}{!}{
    \begin{tabular}{llcccccc}
    \toprule
          & Methods & Original & \Algnameabbr{} $N\!=\!0$ & \Algnameabbr{} $N\!=\!2$ & \Algnameabbr{} $N\!=\!4$ & \Algnameabbr{} $N\!=\!6$ & \Algnameabbr{} $N\!=\!8$ \\
     \midrule
         \multirow{6}{*}{Llama-3-8B} & Latency/token & 0.031 & \cellcolor{lightred} 0.036~(1.16$\times$) & \cellcolor{lightred} 0.068~(2.19$\times$) & \cellcolor{lightgreen} \textbf{0.134~(4.32$\times$)} & \cellcolor{lightgreen} 0.228~(7.35$\times$) & \cellcolor{lightgreen} 0.324~(10.45$\times$) \\
         & TruthfulQA & 0.360 & \cellcolor{lightred} 0.224 & \cellcolor{lightred} 0.354 & \cellcolor{lightgreen} 0.361 & \cellcolor{lightgreen} 0.362 & \cellcolor{lightgreen} 0.362 \\
         & MathQA & 0.421 & \cellcolor{lightred} 0.180 & \cellcolor{lightred} 0.405 & \cellcolor{lightgreen} 0.422 & \cellcolor{lightgreen} 0.421 & \cellcolor{lightgreen} 0.422 \\
         & MMLU & 0.638 & \cellcolor{lightred} 0.252 & \cellcolor{lightgreen} 0.639 & \cellcolor{lightgreen} 0.638 & \cellcolor{lightgreen} 0.638 & \cellcolor{lightgreen} 0.638 \\
         & OpenbookQA & 0.340 & \cellcolor{lightred} 0.172 & \cellcolor{lightgreen} 0.340 & \cellcolor{lightgreen} 0.340 & \cellcolor{lightgreen} 0.342 & \cellcolor{lightgreen} 0.342 \\
         & Average & 0.440 & \cellcolor{lightred} 0.207 & \cellcolor{lightred} 0.434 & \cellcolor{lightgreen} 0.440 & \cellcolor{lightgreen} 0.441 & \cellcolor{lightgreen} 0.441 \\
    \midrule
         \multirow{6}{*}{Mistral-7B} & Latency/token & 0.030 & \cellcolor{lightred} 0.036~(1.2$\times$) & \cellcolor{lightred} 0.070~(2.33$\times$) & \cellcolor{lightred} 0.120~(4$\times$) & \cellcolor{lightred} 0.185~(6.17$\times$) & \cellcolor{lightgreen} \textbf{0.262~(8.73$\times$)} \\
         & TruthfulQA & 0.523 & \cellcolor{lightred} 0.236 & \cellcolor{lightred} 0.503 & \cellcolor{lightred} 0.493 & \cellcolor{lightred} 0.512 & \cellcolor{lightgreen} 0.52 \\
         & MathQA & 0.371 & \cellcolor{lightred} 0.185 & \cellcolor{lightred} 0.347 & \cellcolor{lightred} 0.364 & \cellcolor{lightred} 0.366 & \cellcolor{lightgreen} 0.37 \\
         & MMLU & 0.59 & \cellcolor{lightred} 0.229 & \cellcolor{lightred} 0.583 & \cellcolor{lightgreen} 0.592 & \cellcolor{lightgreen} 0.589 & \cellcolor{lightgreen} 0.588 \\
         & OpenbookQA & 0.36 & \cellcolor{lightred} 0.15 & \cellcolor{lightgreen} 0.362 & \cellcolor{lightgreen} 0.362 & \cellcolor{lightgreen} 0.362 & \cellcolor{lightgreen} 0.36 \\
         & Average & 0.461 & \cellcolor{lightred} 0.2 & \cellcolor{lightred} 0.448 & \cellcolor{lightred} 0.452 & \cellcolor{lightred} 0.457 & \cellcolor{lightgreen} 0.460 \\
    \midrule
         \multirow{6}{*}{Phi-2} & Latency/token & 0.023 & \cellcolor{lightred} 0.026~(1.13$\times$) & \cellcolor{lightred} 0.040~(1.74$\times$) & \cellcolor{lightred} 0.054~(2.35$\times$) & \cellcolor{lightred} 0.072~(3.13$\times$) & \cellcolor{lightgreen} \textbf{0.089~(3.87$\times$)} \\
         & TruthfulQA & 0.306 & \cellcolor{lightred} 0.23 & \cellcolor{lightred} 0.266 & \cellcolor{lightgreen} 0.306 & \cellcolor{lightgreen} 0.306 & \cellcolor{lightgreen} 0.305 \\
         & MathQA & 0.308 & \cellcolor{lightred} 0.203 & \cellcolor{lightred} 0.271 & \cellcolor{lightgreen} 0.308 & \cellcolor{lightgreen} 0.302 & \cellcolor{lightgreen} 0.305 \\
         & MMLU & 0.544 & \cellcolor{lightred} 0.231 & \cellcolor{lightred} 0.406 & \cellcolor{lightred} 0.515 & \cellcolor{lightred} 0.534 & \cellcolor{lightgreen} 0.54 \\
         & OpenbookQA & 0.392 & \cellcolor{lightred} 0.196 & \cellcolor{lightred} 0.35 & \cellcolor{lightred} 0.374 & \cellcolor{lightgreen} 0.388 & \cellcolor{lightgreen} 0.394 \\
         & Average & 0.388 & \cellcolor{lightred} 0.215 & \cellcolor{lightred} 0.323 & \cellcolor{lightred} 0.375 & \cellcolor{lightred} 0.382 & \cellcolor{lightgreen} 0.386 \\
     \bottomrule
    \end{tabular}
    }
\vspace{-4mm}
\end{table*}

\section{\Algnameabbr{} Benefits LLM Community}

\Algnameabbr{} benefits LLM community by protecting the ownership of developers, preventing abuse of LLMs, defense against user fine-tuning, and ensuring secure uses of LLMs.




\subsection{Protecting LLM Ownership}

\Algnameabbr{} secures the LLM weights $\textbf{W}$, $\textbf{b}$, and $\textbf{c}$ by transforming them into $\{\boldsymbol{\Theta}_{i,0}, \cdots, \boldsymbol{\Theta}_{i,N}\}_{i=1}^{D}$, enabling token generations without disclosing their specific values. 
Moreover, theoretically, it is infeasible to precisely derive the original values from the exposed parameters $\{\boldsymbol{\Theta}_{i,0}, \cdots, \boldsymbol{\Theta}_{i,N}\}_{i=1}^{D}$ and $\textbf{z}_0$.
In this way, \Algnameabbr{} preserves the ownership of developers on their released LLMs.

\subsection{Preventing Abuse of LLMs by Low-speed Token Generation}

\Algnameabbr{} induces low-speed token generation for the secured LLMs by incorporating approximately $N \times$ the floating point operations~(FLOPs) compared with the original MLPs.
The FLOPs of Equation~(\ref{eq:talor_forward}) are $N \times$ those of Equation~(\ref{eq:mlp_inference}). 
This can significantly reduce the token generation speed of unauthorized use, making such usage low-utility for unauthorized users.
This intentional delay helps developers prevent the potential large-scale unauthorized use of their released LLMs.
We note this intentional delay as the ``Taylor Unswift".



\subsection{Defense against LoRA-based Fine-tuning Methods}

LoRA~\cite{hu2021lora} is a powerful method for fine-tuning LLMs on user-specific datasets.
However, it is necessary for the LoRA-based methods to have the original LLM weights for the inference process.
Therefore, \Algnameabbr{} can naturally defend against LoRA-based fine-tuning by not exposing the original LLM weights.

\subsection{Harness of LLMs' Capability without Data Privacy Concerns}

\Algnameabbr{} allows unauthorized users to run and test LLMs on private datasets before applying for authorization, as shown in Figure~\ref{fig:LLM_access}~(c). This testing process can be fully conducted by the users~(data owners) without sharing their private data with the developers. Based on the testing results on their private data, users can decide whether or not to apply for authorization.



\subsection{Ensuring Secure Use of LLMs.}

\Algnameabbr{} facilitates large-scale applications of LLMs under regulatory or contractual constraints.
Specifically, \Algnameabbr{} increases $N \times$ latency of token generation, which cannot meet the efficiency requirements for large-scale applications. 
As a result, users need to request authorization from developers for the LLM weight values. 
This process enables developers to address security concerns through authorization regulations or contracts, ensuring the secure use of the LLMs.
In this way, large-scale applications on the user side are under the constraints of regulations or contracts.

\section{Experiments}

In this section, we conduct experiments to evaluate \Algnameabbr{} by answering the following research questions:
\textbf{RQ1:} Can \Algnameabbr{} retain the accuracy of original LLMs while adjusting generation speed?
\textbf{RQ2:} How does \Algnameabbr{} defend against fine-tuning on downstream datasets and distilling on large-scale datasets?
\textbf{RQ3:} How does the expansion order influence the output of \Algnameabbr{} compared with that of original LLMs?

\begin{table*}[]
\centering
\caption{Accuracies of the original LLMs, \Algnameabbr{}, and fine-tuned LLMs on downstream tasks.}
\vspace{-2mm}
    \label{tab:ft_defense}  
    \small
    \begin{tabular}{llcccccc}
    \toprule
         Datasets & & TruthfulQA & MathQA & MMLU & OpenbookQA & Average \\
     \midrule
         \multirow{3}{*}{Mistral-7B} & Original & 0.523 & 0.371 & 0.590 & 0.360 & 0.461 \\ 
         & \Algnameabbr{} $N=8$ & 0.520 & 0.370 & 0.588 & 0.360 & \textbf{0.460} \\
         & Fine-tuning & 0.090 & 0.080 & 0.120 & 0.020 & 0.078 \\
    \midrule
         \multirow{3}{*}{Phi-2} & Original & 0.306 & 0.308 & 0.544 & 0.392 & 0.388 \\ 
         & \Algnameabbr{} $N=8$ & 0.305 & 0.305 & 0.540 & 0.394 & \textbf{0.386} \\
         & Fine-tuning & 0.085 & 0.107 & 0.068 & 0.130 & 0.098 \\
     \bottomrule
    \end{tabular}
\end{table*}

\subsection{Experiment Setup}

We specify the datasets, LLMs, evaluation metrics, and implementation details.

\paragraph{Datasets.}
The evaluation of \Algnameabbr{}~is based on the TruthfulQA~\cite{lin2021truthfulqa}, MathQA~\cite{amini-etal-2019-mathqa}, MMLU~\cite{hendryckstest2021}, and OpenbookQA~\cite{OpenBookQA2018} datasets.
We use the \texttt{lm-evaluation-harness}~\cite{gao10256836framework} as the codebase for the experiments of evaluation.


\paragraph{LLMs.} We evaluate \Algnameabbr{} using three popular model families: Llama-3-8B~\cite{touvron2023llama}, Mistral-7B~\cite{jiang2024mixtral}, and Phi-2~\cite{textbooks2}.
We download these models from the Huggingface Transformers~\cite{wolf2019huggingface}.

\paragraph{Evaluation Metrics.} \!\!\!\!\!\!
We evaluate the \texttt{accuracy}($\uparrow$) of LLMs on downstream datasets to determine whether \Algnameabbr{} can preserve the accuracy of original LLMs.
Moreover, we measure the \texttt{per-token latency} to assess the generation speed \cite{liu2024kivi}.
It is the time cost of generating a single token.
To determine if \Algnameabbr{}'s outputs align with original LLM's outputs, we measure the \texttt{Kullback}\texttt{–Leibler} \texttt{divergence}($\downarrow$) and \texttt{ROUGE-1 score}($\uparrow$) of \Algnameabbr{}'s outputs, using the original LLM's outputs as the ground-truth values.
We also include case studies for evaluating if \Algnameabbr{} preserves the chat capability of LLMs.

\paragraph{Implementation Details.} 
\Algnameabbr{} protects the $\mathrm{d}_{\text{model}} \!\times \! \mathrm{d}_{\text{intermediate}}$ \texttt{down\_projection}~weights within each layer of LLMs, as shown in Figure~\ref{fig:protect_mlp}~(b).
Our empirical studies show that securing an $\mathrm{d}_{\text{model}} \times M$ submatrix of the \texttt{down\_projection} weights is sufficient for protection, where $\mathrm{d}_{\text{model}} \! \leq \! M \! \leq \! \mathrm{d}_{\text{intermediate}}$;
Specifically, for the Llama-3-8B, Mistral-7B, and Phi-2 LLMs, their $\mathrm{d}_{\text{model}}$ are 4096, 4096, and 2560, while their $\mathrm{d}_{\text{intermediate}}$ values are 14336, 14336, and 10240, respectively. 
Therefore, \Algnameabbr{} targets 1$\times 10^4$, 8$\times 10^3$, and 2560 rows of the \texttt{down\_projection} weights for the Llama-3-8B, Mistral-7B, and Phi-2 LLMs, respectively.
Given that these LLMs consist of 32 layers, \Algnameabbr{} actually protects 1.31B, 1.05B, and 210M parameters for Llama-3-8B, Mistral-7B, and Phi-2, respectively. 
More details are in Appendix~\ref{appendix:implement_details}.

\begin{figure*}
\footnotesize
\fcolorbox{black}{lightgray!15}{\parbox{.99\linewidth}
{
\textbf{Context:} 
Although initially he was little @-@ known to other writers , his works came to be hugely influential in both Chinese and Japanese literary culture . $\cdots$ while the range of his work has allowed him to be introduced to Western readers as ' the Chinese Virgil , Horace , \textcolor{darkred}{<unk>} , Shakespeare , Milton , Burns , \textcolor{darkred}{<unk>} , \textcolor{darkred}{<unk>}.
\vspace{1mm}
\\
\textbf{Original Answer:} 
It seems like there are some missing words in the text. Based on the context, I'm going to try to fill in the gaps: Although initially he was little known to other writers, his works came to be hugely influential in both Chinese and Japanese literary culture. $\cdots$ while the range of his work has allowed him to be introduced to Western readers as ' the Chinese Virgil , Horace, \textcolor{darkgreen}{Dante}, Shakespeare, Milton, Burns, \textcolor{darkgreen}{Goethe}, and \textcolor{darkgreen}{Hugo}'.
\vspace{1mm}
\\
\textbf{Distilled LLM's Answer:} \textcolor{darkred}{addCriterion addCriterion addCriterion addCriterion ... addCriterion}
\vspace{1mm}
\\
\textbf{\Algnameabbr{} $N=8$ Answer:} 
It seems like there are some missing words in the text. Based on the context, I'm going to try to fill in the gaps: Although initially he was little known to other writers, his works came to be hugely influential in both Chinese and Japanese literary culture. $\cdots$ while the range of his work has allowed him to be introduced to Western readers as 'the Chinese Virgil, Horace, \textcolor{darkgreen}{Dante}, Shakespeare, Milton, Burns, \textcolor{darkgreen}{Goethe}, and \textcolor{darkgreen}{Hugo}'.
}}
\vspace{-2mm}
\caption{\label{fig:case_study2} \small The outputs of original LLMs, distilled LLMs, and \Algnameabbr{}. The input context is from the wikitext-2 dataset.}
\vspace{-2mm}
\end{figure*}

\subsection{Accuracy and Latency Analysis~(RQ1)}

The accuracy and per-token latency of \Algnameabbr{} under different expansion orders are illustrated in Table~\ref{tab:acc_latency}.
These results are compared with the performance and latency of original Llama-3-8B, Mistral-7B, and Phi-2 LLMs.


\paragraph{Retaining Accuracy.}
According to Table~\ref{tab:acc_latency}, for the Llama-3-8B, Mistral-7B, and Phi-2 LLMs, \Algnameabbr{} with $N = \text{4, 8}$, and 8 are generally as accurate as the original LLMs across all datasets.


\paragraph{Adjusting Generation Speed.}
For Llama-3-8B, Mistral-7B, and Phi-2 LLMs, we focus on the per-token latency of $N = \text{4, 8}$, and 8, where \Algnameabbr{} can generally retain the accuracy of original LLMs.
Notably, \Algnameabbr{} has \textbf{4.32$\times$}, \textbf{8.73$\times$} and \textbf{3.73$\times$} increase of the latency compared with the original LLMs, respectively.
By reducing the generation speed, \Algnameabbr{} potentially prevents large-scale unauthorized use of released LLMs.


\paragraph{Agnostic to Different LLMs.}
\Algnameabbr{} is a model agnostic method to protect the ownership of LLMs.
Applied to different LLMs, it shows consistent capacity in retaining the performance, while inducing the low-speed generation process. 


\subsection{Defending against Fine-tuning~(RQ2)}
\label{sec:exp_ft}

We demonstrate the capability of \Algnameabbr{} in defending a reconstruction of the secured weights by dataset-based fine-tuning.
Specifically, we play as unauthorized users to randomly reinitialize the secured weights $\textbf{b}$, $\textbf{W}$, and $\textbf{c}$ of the MLP layers within the LLMs and attempt to reconstruct their values through fine-tuning processes on labeled datasets. 
The fine-tuning processes are conducted using the Mistral-7B and Phi-2 LLMs on the four downstream datasets: TruthfulQA, MathQA, MMLU, and OpenbookQA. 
The hyperparameters settings are provided in Appendix~\ref{appendix:ft_hyper}.
The fine-tuning results are compared with those of original LLMs and \Algnameabbr{} with $N=8$ in Table~\ref{tab:ft_defense}.


\paragraph{Accuracy.} 
According to Table~\ref{tab:ft_defense}, the fine-tuned LLMs exhibit a significant decline in accuracy across each downstream task when compared to both the original LLMs and \Algnameabbr{}. Learning such a vast number of parameters from scratch (1.05B for the Mistral-7B and 210M for Phi-2) is highly challenging, given the limited instances and label information available from downstream datasets. 
This challenge arises because \Algnameabbr{} secures the pre-trained weights of LLMs, preventing effective initialization for fine-tuning.
This indicates the effectiveness of \Algnameabbr{} in defending unauthorized users from using downstream datasets to reconstruct the protected weights.


\subsection{Defending against Distillation~(RQ2)}
\label{sec:exp_distill}

We demonstrate the capability of \Algnameabbr{} in defending a reconstruction of the secured weights by knowledge distillation. 
Specifically, we play as unauthorized users to randomly reinitialize the secured weights $\textbf{b}$, $\textbf{W}$, and $\textbf{c}$ of the MLP layers within the LLMs and attempt to reconstruct their values by learning the output distribution of the original LLMs.
The distillation process is conducted on the C4-En dataset~\cite{raffel2020exploring}, and the distilled LLMs are evaluated on the WikiText-2 dataset~\cite{merity2016pointer} using the perplexity metric~($\downarrow$). We also compare the perplexity of the original LLMs and \Algnameabbr{} with $N=8$. The experiment is conducted on the Llama-3-8B LLM due to its state-of-the-art capabilities. The hyperparameter settings are provided in Appendix~\ref{appendix:ft_hyper}.



\paragraph{Perplexity.}
The distilled LLMs exhibit considerably higher perplexity~(265.62) compared to the original LLMs~(12.72), as shown in Table~\ref{tab:perplexity_distillation}.
In contrast, \Algnameabbr{} with $N=8$ achieves a comparable perplexity~(12.75) to the original result~(12.72). 
This demonstrates that, without access to the protected weights, unauthorized users cannot use distillation methods to match the language modeling performance of the original LLMs.

\paragraph{Hallucination of Distilled LLMs.}
According to Figure~\ref{fig:case_study2}, when given contexts from the Wikitext-2 dataset, the distilled LLMs generate "addCriterion" that are complete hallucinations. 
In contrast, \Algnameabbr{} with $N=8$ produces tokens that align precisely with those of the original LLMs, delivering coherent and accurate answers. This indicates that unauthorized users cannot fully restore the chat capabilities of the LLMs by reinitializing the protected weights and distilling the LLMs.


\begin{table}
    \centering
    \caption{\small Perplexity on the WikiText-2 dataset.}
    \vspace{-2mm}
    \label{tab:perplexity_distillation}
    \resizebox{0.49\textwidth}{!}{
    \begin{tabular}{lccc}
    \toprule
         Method & Original & \Algnameabbr{} $N\!=\!8$ & Distilled LLMs \\
    \midrule
         Perplexity & 12.72 & \textbf{12.75} & 256.62 \\
     \bottomrule
    \end{tabular}
    }
\vspace{-3mm}
\end{table}

\begin{figure*}[t!]
\footnotesize
\fcolorbox{black}{lightgray!15}{\parbox{.99\linewidth}
{
\textbf{Context:} New York (CNN) -- More than 80 Michael Jackson collectibles -- including the late pop star's famous rhinestone-studded glove from a 1983 performance -- were auctioned off Saturday, reaping a total $\$$2 million. Profits from the auction at the Hard Rock Cafe in New York's Times Square crushed pre-sale expectations of only $\$$120,000 in sales. $\cdots$ Hoffman Ma, who bought the glove on behalf of Ponte 16 Resort in Macau, paid a 25 percent buyer's premium, which was tacked onto all final sales over $\$$50,000. Winners of items less than $\$$50,000 paid a 20 percent premium.
\vspace{1mm}
\\
\textbf{Question:} Where was the Auction held? How much did they make? How much did they expected? Who buy the Jackson Glove Where was the buyer of the glove from?
\vspace{1mm}
\\
\textbf{Original Answer:} According to the article, the auction was held at the Hard Rock Cafe in New York's Times Square. The auction made a total of $\$$2 million, which is more than the expected $\$$120,000. The Michael Jackson glove was sold for $\$$420,000 to a buyer from Hong Kong, China, who was representing Ponte 16 Resort in Macau.
\vspace{1mm}
\\
\textbf{\Algnameabbr{} $N=0$ Answer:} \textcolor{darkred}{\_REFascus Bout Bout Bout Bout Bout Bout Bout Bout $\cdots$ Bout}
\vspace{1mm}
\\
\textbf{\Algnameabbr{} $N=1$ Answer:} \textcolor{darkred}{The article was written from the article from $\cdots$ from the article}
\vspace{1mm}
\\
\textbf{\Algnameabbr{} $N=2$ Answer:} Here are the answers to your questions: 
\textcolor{darkred}{1. Where was the Auction held?} \textcolor{darkgreen}{The auction was held at the Hard Rock Cafe in New York's Times Square.} 
\textcolor{darkred}{2. How much did they make?} 
\textcolor{darkgreen}{The auction raised a total of $\$$2 million.} 
\textcolor{darkred}{3. How much did they expect?} 
\textcolor{darkgreen}{The auction house, Julien's Auctions, expected the auction to raise only $\$$120,000.} 
\textcolor{darkred}{4. Who bought the Jackson Glove?} 
\textcolor{darkgreen}{The Jackson glove was bought by Hoffman Ma, who purchased it on behalf of Ponte 16 Resort in Macau.} 
\textcolor{darkred}{5. Where was the buyer of the glove from?} 
\textcolor{darkgreen}{The buyer of the glove, Hoffman Ma, is from Hong Kong, China.}
\vspace{1mm}
\\
\textbf{\Algnameabbr{} $N=8$ Answer:} \textcolor{darkgreen}{According to the article, the auction was held at the Hard Rock Cafe in New York's Times Square. The auction made a total of $\$$2 million, which is more than the expected $\$$120,000. The Michael Jackson's famous rhinestone-studded glove from a 1983 performance was sold for $\$$420,000 to a buyer from Hong Kong, China, who is Hoffman Ma, representing Ponte 16 Resort in Macau.}
}}
\caption{\label{fig:case_study1} The input context and question from the CoQA dataset; the outputs of original LLMs; and the outputs of \Algnameabbr{} whose the expansion order $N$ takes 0, 1, 2, and 8.}
\vspace{-3mm}
\end{figure*}

\begin{figure}
\centering
\includegraphics[width=0.95\linewidth]{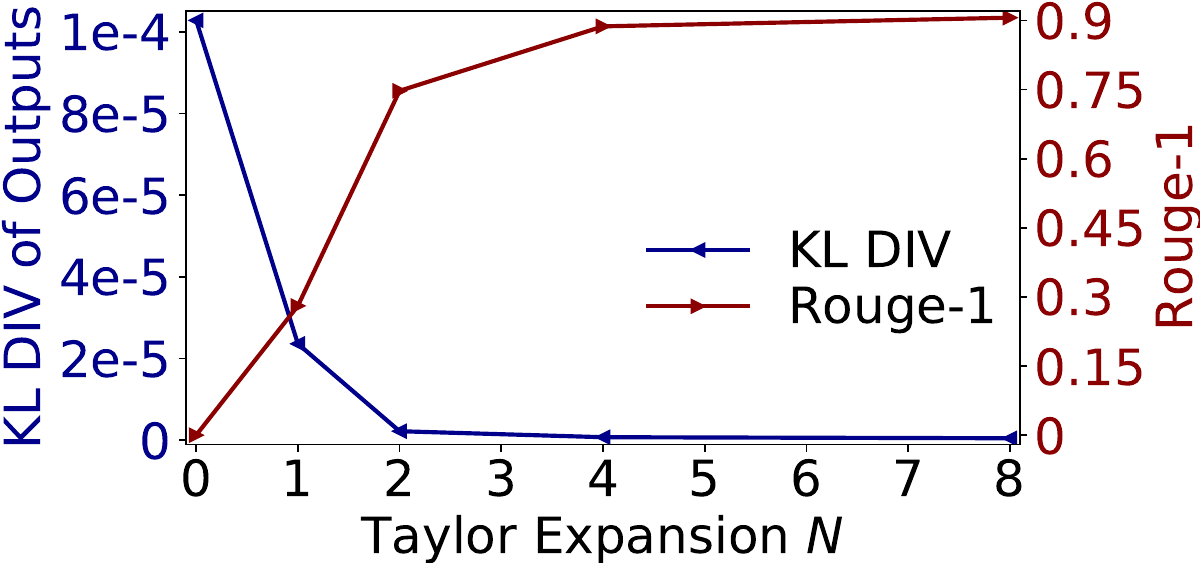}
\vspace{-2mm}
\caption{\label{fig:kl_vs_n} The KL divergence of output probability and ROUGE-1 score versus the expansion order.}
\vspace{-3mm}
\end{figure}

\subsection{Influence of Expansion Order for \Algnameabbr{}~(RQ3)}
\label{sec:exp_expan_order}

We study the influence of expansion order on the outputs of \Algnameabbr{}. Specifically, the outputs of the original LLMs are taken as ground-truth values, and those of \Algnameabbr{} with different expansion orders are compared with these ground-truth values. We employ the Kullback–Leibler divergence~($\downarrow$) and ROUGE-1~($\uparrow$) to quantitatively measure the distance between the outputs of \Algnameabbr{} and the ground-truth values, which represent statistical and token space distance, respectively. These experiments are conducted on the CoQA dataset~\cite{roemmele2011choice}. The experimental results are shown in Figure~\ref{fig:kl_vs_n}. Additionally, we also show the generated tokens from \Algnameabbr{} with different expansion orders in Figure~\ref{fig:case_study1}.




\paragraph{\Algnameabbr{}'s Outputs Gradually Converge to Original LLMs.}
According to Figure~\ref{fig:kl_vs_n}, as $N$ grows from 0 to 8, the Kullback–Leibler divergence decreases to zero, while the ROUGE-1 score rises to 0.9. This trend is consistent with our theoretical discussion in Section~\ref{sec:theory}.
When expansion order $N \geq 8$ is sufficiently large, the outputs of \Algnameabbr{} closely match those of the original LLMs.

\paragraph{Hallucination Caused by Insufficient Taylor Expansion Order.}
According to Figure~\ref{fig:case_study1}, given the context and question from the CoQA dataset, \Algnameabbr{} with $N=0$ or $N=1$ output "Bout" or "the article was written from," which are complete hallucinations. With $N=2$, \Algnameabbr{} provides the key points of the answers but undesirably repeats the questions. An expansion order of $N=8$ is sufficient for \Algnameabbr{}, as the output context closely matches that of original LLMs.

\paragraph{Effectiveness.}
According to both the quantitative and qualitative results, a sufficiently large expansion order, such as $N \geq 8$, is necessary for unauthorized users to avoid the hallucinations.
In this way, \Algnameabbr{} adjusts the unauthorized generation process $4\times$ slower than original LLMs.



\section{Related Work}

\paragraph{File Encryption.} Encryption technologies can effectively prevent the unauthorized use of digital files, including the checkpoint file of large language models (LLMs). Representative algorithms include the Advanced Encryption Standard~(AES)~\cite{bogdanov2011biclique}, Blowfish~\cite{rijmen1997cryptanalysis}, and Rivest–Shamir–Adleman~(RSA)~\cite{rivest1978method}. However, these methods are not suitable for LLMs because the models cannot perform inference using encrypted weight values. This limitation prevents users from harnessing the capabilities of released LLMs on their private datasets, which may cause a loss of market opportunities.

\paragraph{Offsite-Tuning.} The offsite-tuning method allows developers to share parts of large language models (LLMs) with users~\cite{xiao2023offsite}. However, this method has several disadvantages. First, users need to perform fine-tuning, which is extremely costly and time-consuming. Additionally, fine-tuning LLMs is challenging for individuals who are not machine learning engineers. These factors hinder the application of offsite-tuning methods in real-world scenarios.

\paragraph{Key Prompt Protection.} The Key Prompt Protection (KPP) aims to prevent the unauthorized use of LLMs~\cite{tangsecure}. With KPP, LLMs respond only when presented with the correct key prompt; otherwise, they will ignore any input instructions. However, a drawback of KPP is that it relies on fine-tuning to embed the protection key into the LLMs. As a result, the LLMs lose their foundational effectiveness and become specialized only in the domains of the fine-tuned datasets.
KPP cannot address the access dilemma because developers cannot access users' private data to embed the protection key into their released LLMs.


\paragraph{Advantages of \Algnameabbr{} over Related Work.}
Unlike file encryption, \Algnameabbr{} allows users to harness the capability of released LLMs on their private datasets while maintaining the protection of model ownership, thus attracting users to apply for authorization. 
Different from offsite-tuning, \Algnameabbr{} enables immediate use of LLMs without the need for fine-tuning.
Moreover, compared to key prompt protection, \Algnameabbr{} preserves the general capabilities of LLMs as foundational models, without restricting them to specific domains.
To summarize, \Algnameabbr{} stands out as a highly effective method for protecting ownership and ensuring secure uses of LLMs.

\paragraph{LLM Watermarks.} Watermark technologies enable developers to detect whether their models have been misused by embedding a digital signature into the authorized model. This signature acts as an identifier that can be detected in misuse scenarios. For API-accessed LLMs and open-sourced LLMs, the signatures are embedded into the distribution of output tokens~\cite{xiang2021protecting} and model weights~\cite{xu2024instructional}, respectively, without compromising the performance of the LLMs. Different from watermarks, TaylorMLP protects the model before it is authorized for weight access by releasing the Taylor-series parameters. It protects the original weight values and allows users to test the models without risking data privacy. TaylorMLP and watermark technologies can significantly complement each other, ensuring the security of models throughout the entire process.

\paragraph{Other Work.} TransLinkGuard~\cite{li2024translinkguard} is a plug-and-play model protection approach against model stealing on edge devices.
MPCFormer~\cite{li2022mpcformer} uses Secure Multi-Party Computation and Knowledge Distillation (KD) for privacy-preserving Transformer inference.

\section{Conclusion}

In this work, we propose \Algnameabbr{} to preserve the ownership of released LLMs and prevent their abuse.
Specifically, \Algnameabbr{} preserves the ownership of LLM by transforming the weights of LLMs into parameters of Taylor-series. Instead of releasing the original weights, developers can release the Taylor-series parameters with users, thereby ensuring the security of LLMs. 
Defensive experiments confirm that \Algnameabbr{} effectively prevents users from reconstructing the weight values based on downstream datasets.
Moreover, \Algnameabbr{} prevents abuse of LLMs by inducing low-speed token generation for the protected LLMs. This intentional delay prevents the potential large-scale unauthorized abuse.
Empirical and case studies show that \Algnameabbr{} significantly increases the latency of token generation, while maintaining the chat capabilities and performance on downstream tasks.
Both qualitative and quantitative results demonstrate the effectiveness of \Algnameabbr{} in ensuring the security of the released LLMs.
This indicates its potential in real-world applications.





\section{Limitations and Potential Risks}
\label{sec:limitations}

In this work, we propose a framework to protect the ownership of LLMs released.
\Algnameabbr{} can facilitate large-scale applications of LLMs under regulatory or contractual constraints.
Upon authorizing weights to users, developers can deliver regulations or contracts to ensure LLM applications comply with specified constraints. 
After authorization, watermark technologies are required as a complementary to detect whether users have violated regulations by misusing or sharing the models with others.
\Algnameabbr{} and watermark complement each other to ensure the security of models throughout the entire process.


\section*{Acknowledgments}

This research was supported by NSF Awards IIS-2224843, and the US Department of Transportation (USDOT) Tier-1 University Transportation Center (UTC) Transportation Cybersecurity Center for Advanced Research and Education (CYBER-CARE) grant \#69A3552348332. \!\!
Additionally, this research was also supported, in part, by NSF Awards OAC-2112606 and OAC-2117439. This work made use of the High Performance Computing Resource in the Core Facility for Advanced Research Computing at Case Western Reserve University (CWRU). We give our special thanks to the CWRU HPC team for their timely and professional help and maintenance. The views and conclusions in this paper are those of the authors and do not represent the views of any funding or supporting agencies.

\bibliography{citation}

\clearpage
\appendix

\section*{Appendix}
\label{sec:appendix}


\section{High-order Derivative of $\text{GELU}(\bullet)$}
\label{appendix:gelu_grad}


\noindent
The $\text{gelu}(\bullet)$ function is given by
\begin{equation}
    \text{gelu}(x) = x \Phi (x),
    \nonumber
\end{equation}
where $\Phi (x)$ denotes the standard Gaussian cumulative distribution function.

\begin{figure*}
\centering
\subfigure[$y=\text{gelu}(x)$]{
\centering
	\begin{minipage}[t]{0.3\linewidth}
		\includegraphics[width=0.99\linewidth]{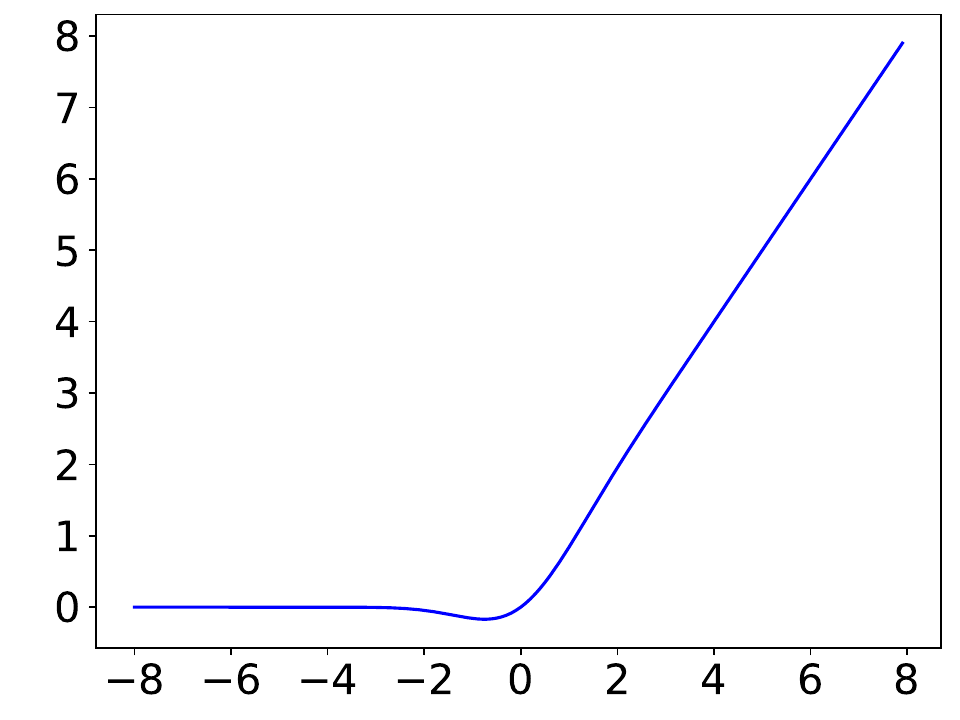}
	\end{minipage}%
}
\subfigure[$y=\text{gelu}^{(1)}(x)$]{
\centering
	\begin{minipage}[t]{0.3\linewidth}
		\includegraphics[width=0.99\linewidth]{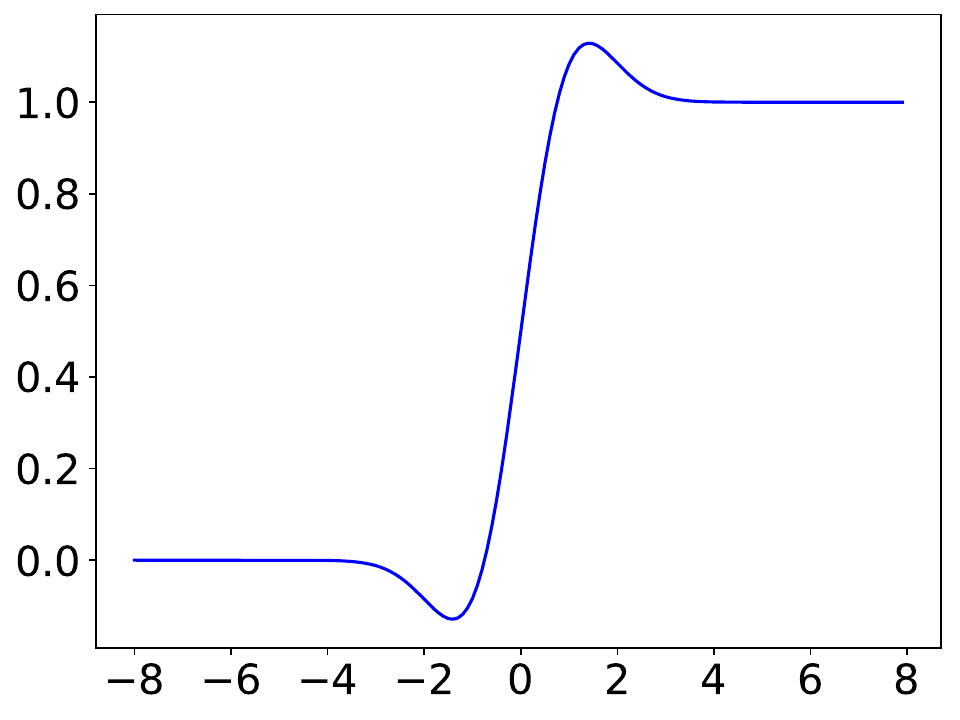}
	\end{minipage}
}
\subfigure[$y=\text{gelu}^{(2)}(x)$]{
\centering
	\begin{minipage}[t]{0.3\linewidth}
		\includegraphics[width=0.99\linewidth]{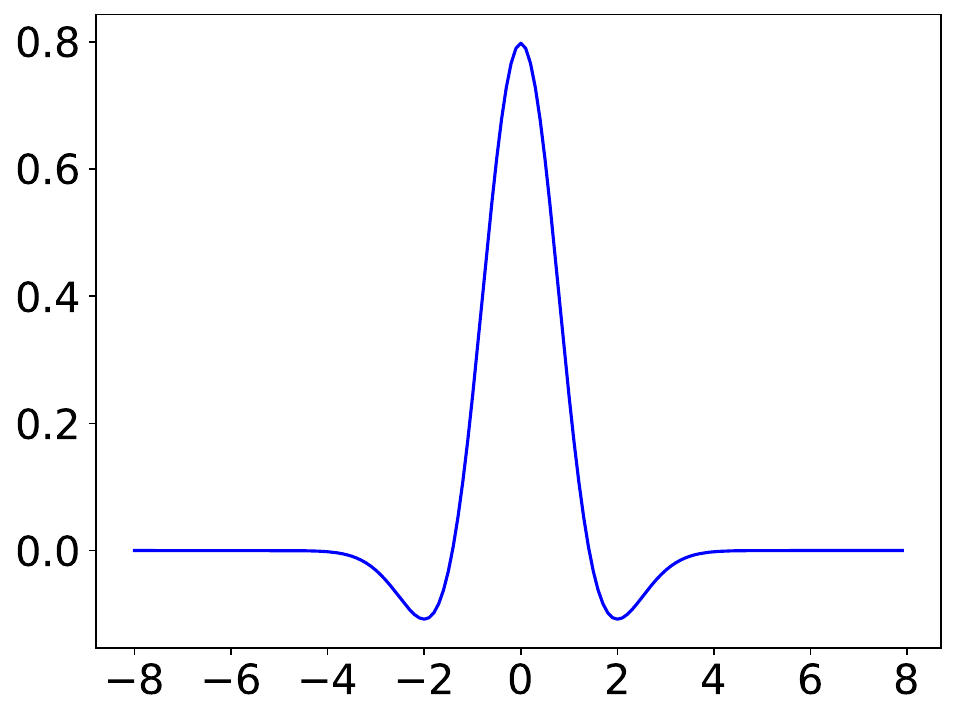}
	\end{minipage}%
}
\subfigure[$y=\text{gelu}^{(3)}(x)$]{
\centering
	\begin{minipage}[t]{0.3\linewidth}
		\includegraphics[width=0.99\linewidth]{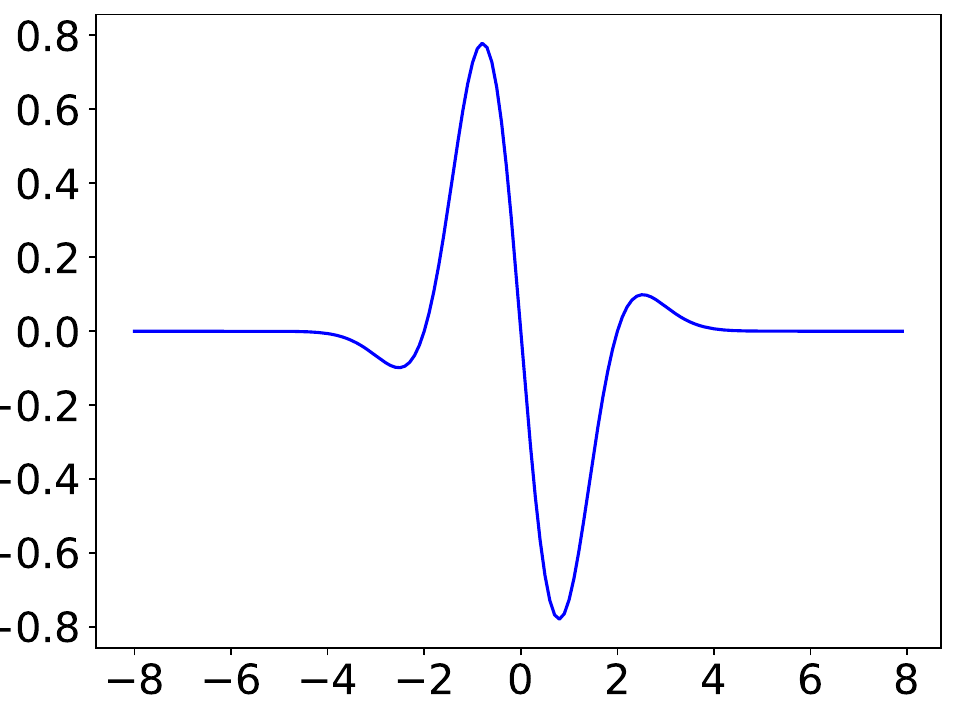}
	\end{minipage}%
}
\subfigure[$y=\text{gelu}^{(4)}(x)$]{
\centering
	\begin{minipage}[t]{0.3\linewidth}
		\includegraphics[width=0.99\linewidth]{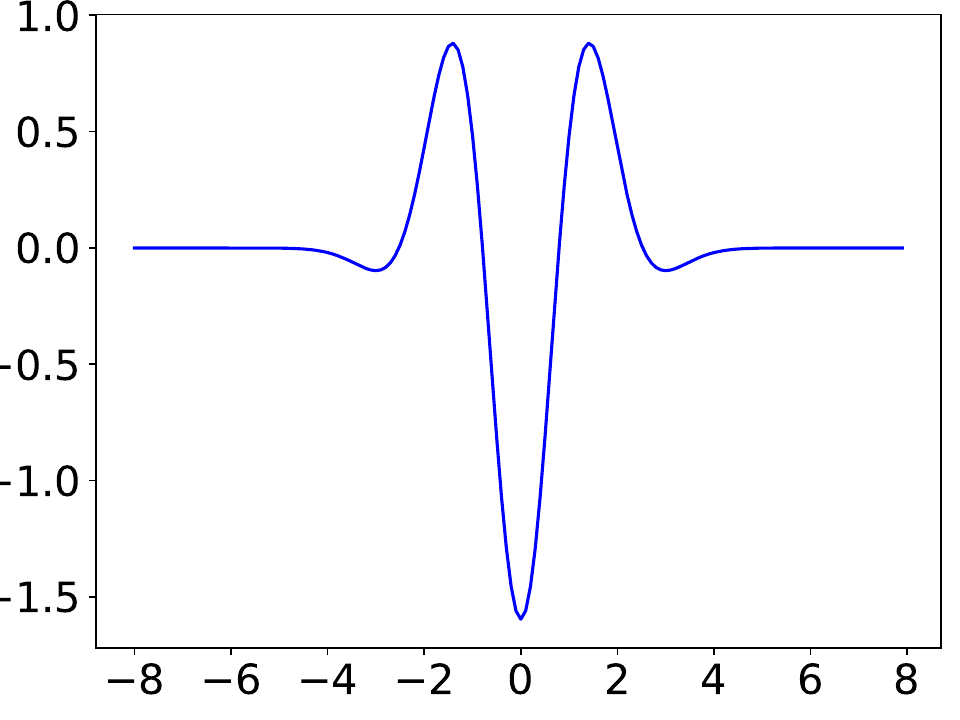}
	\end{minipage}
}
\subfigure[$y=\text{gelu}^{(5)}(x)$]{
\centering
	\begin{minipage}[t]{0.3\linewidth}
		\includegraphics[width=0.99\linewidth]{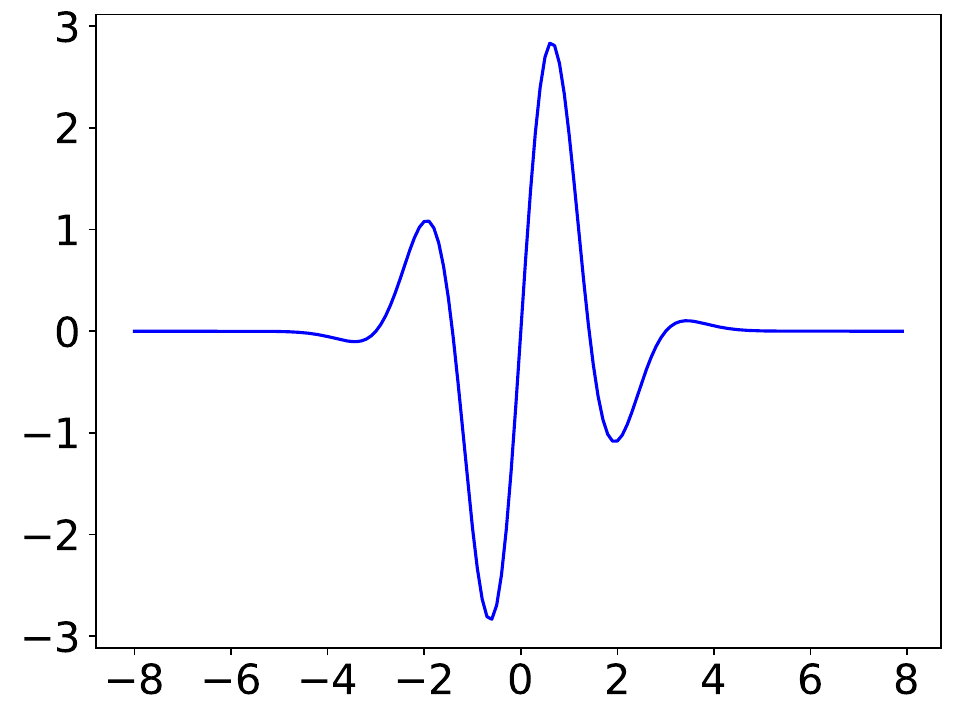}
	\end{minipage}%
}
\caption{Visualization of $y=\text{gelu}^{(n)}(x)$, where $n = 0,1,2,3,4,5$.}
\label{fig:gelu_grad}
\end{figure*}

\noindent
According to the Leibniz rule, we have
\begin{align}
    \text{gelu}^{(n)}(x) &= \sum_{k=0}^n \binom{k}{n} x^{(k)} \Phi^{(n-k)} (x)
    \nonumber
    \\
    &= x \Phi^{(n)} (x) + n \Phi^{(n-1)} (x).
\end{align}
We let
\begin{align}
    &h_n (x) = \frac{1}{\sqrt{2 \pi}} \Big( e^{-\frac{x^2}{2}} \Big)^{(n)}
    \nonumber
    \\
    &= \! \Big[ (-x) \frac{1}{\sqrt{2 \pi}} \Big( \! e^{-\frac{x^2}{2}} \! \Big)^{(n-1)} \!\!\!\!\!\!\!\!\!\!\!\! -\! (n\!-\!1) \frac{1}{\sqrt{2 \pi}} \Big( \! e^{-\frac{x^2}{2}} \! \Big)^{(n-2)} 
    \nonumber
    \\
    \label{eq:gelu_h_n}
    &= (-x) h_{n-1} - (n-1) h_{n-2}.
\end{align}
The $n$-order derivative of $\Phi^{(n)} (x)$ is given by
\begin{align}
    \Phi^{(n)} (x) &= \frac{1}{\sqrt{2 \pi}} \Big( e^{-\frac{x^2}{2}} \Big)^{(n-1)} =  h_{n-1} (x)
    \nonumber
    \\
    \Phi^{(n-1)} (x) &= \frac{1}{\sqrt{2 \pi}} \Big( e^{-\frac{x^2}{2}} \Big)^{(n-2)} =  h_{n-2} (x).
    \nonumber
\end{align}
The $n$-order derivative of $\text{gelu}(\bullet)$ is given by
\begin{equation}
    \text{gelu}^{(n)}(x) = x h_{n-1}(x) + n h_{n-2}(x),
\end{equation}
where $h_n (x)$ can be recursively computed by Equation~(\ref{eq:gelu_h_n}); $h_0(x) = \frac{1}{\sqrt{2 \pi}} e^{-\frac{x^2}{2}}$; and $h_1(x) = -x h_0(x)$.
%
We also visualize $y=\text{gelu}^{(n)}(x)$ with $n = 0,1,2,3,4,5$ in Figure~\ref{fig:gelu_grad}~(a)-(f), respectively.




\section{High-order Derivative of $\text{SiLU}(\bullet)$}
\label{appendix:silu_grad}

The $\text{silu}(\bullet)$ function is given by
\begin{equation}
    \text{silu}(x) = x \sigma (x),
    \nonumber
\end{equation}
where $\sigma(x)$ denotes the sigmoid function.

\begin{figure*}
\centering
\subfigure[$y=\text{silu}(x)$]{
\centering
	\begin{minipage}[t]{0.3\linewidth}
		\includegraphics[width=0.99\linewidth]{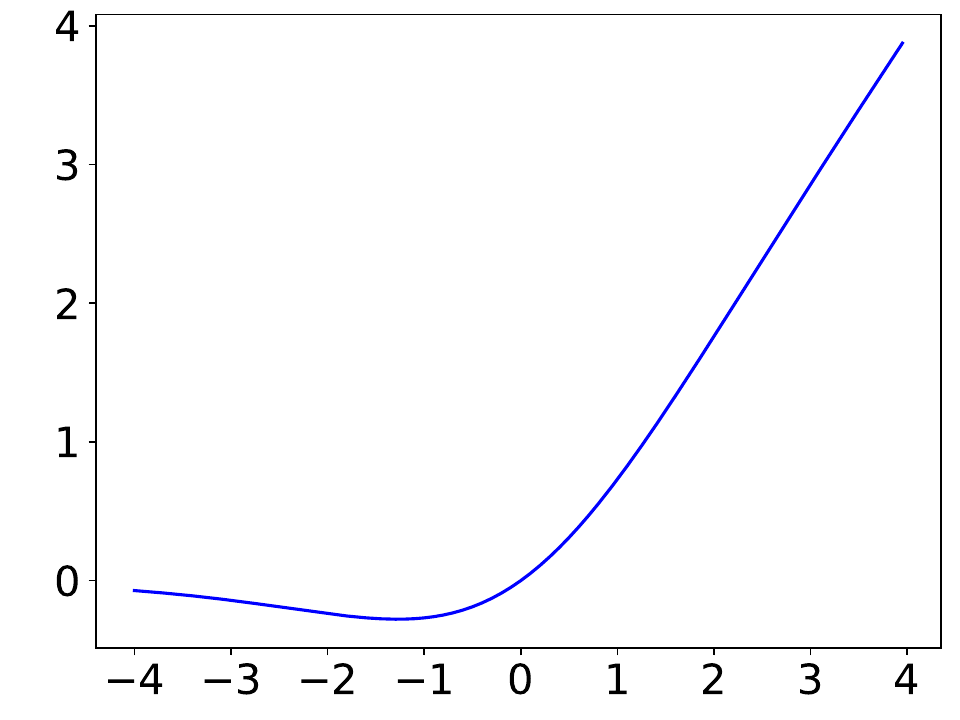}
	\end{minipage}%
}
\subfigure[$y=\text{silu}^{(1)}(x)$]{
\centering
	\begin{minipage}[t]{0.3\linewidth}
		\includegraphics[width=0.99\linewidth]{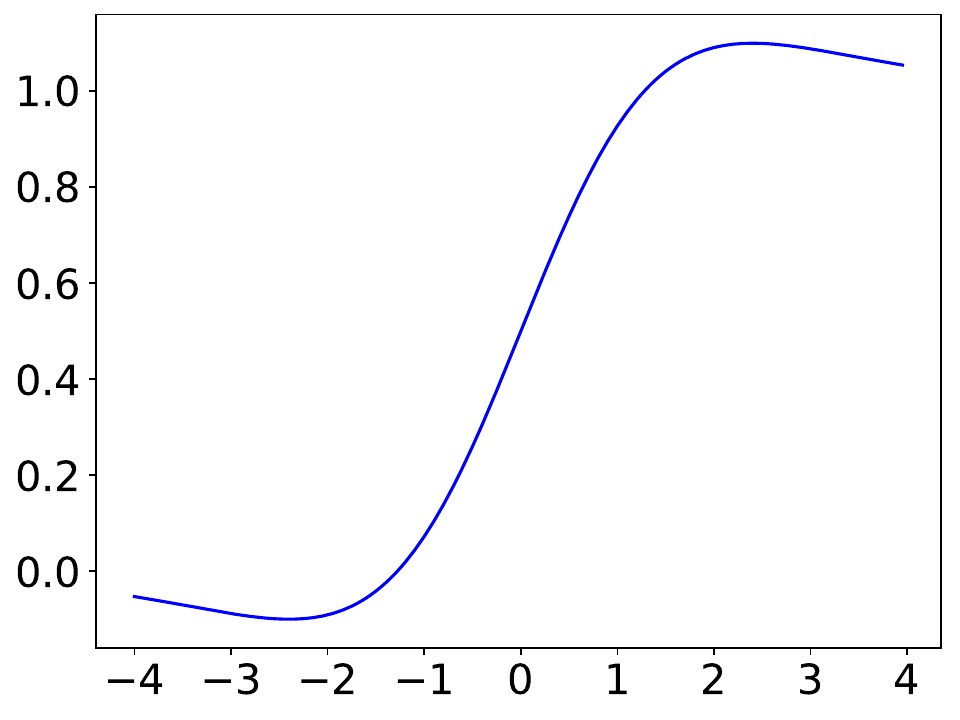}
	\end{minipage}
}
\subfigure[$y=\text{silu}^{(2)}(x)$]{
\centering
	\begin{minipage}[t]{0.3\linewidth}
		\includegraphics[width=0.99\linewidth]{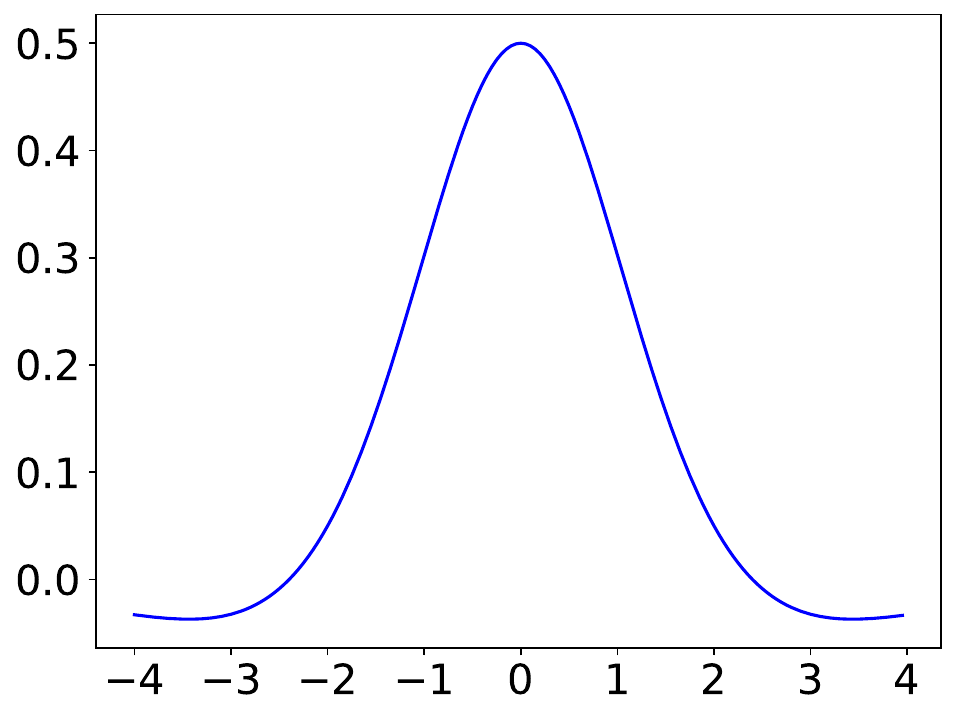}
	\end{minipage}%
}
\subfigure[$y=\text{silu}^{(3)}(x)$]{
\centering
	\begin{minipage}[t]{0.3\linewidth}
		\includegraphics[width=0.99\linewidth]{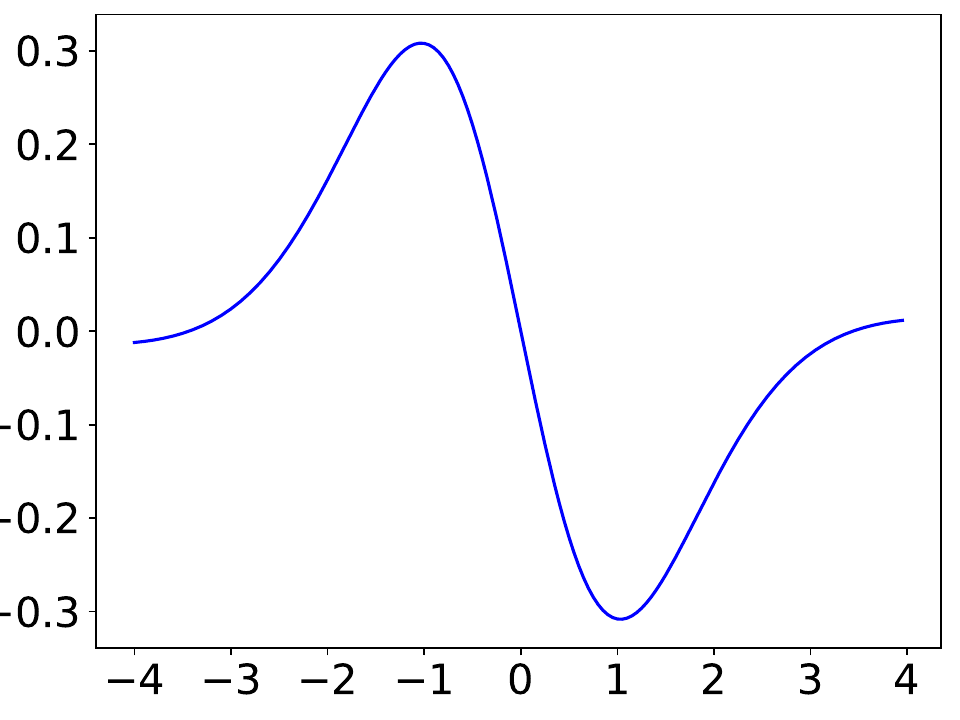}
	\end{minipage}%
}
\subfigure[$y=\text{silu}^{(4)}(x)$]{
\centering
	\begin{minipage}[t]{0.3\linewidth}
		\includegraphics[width=0.99\linewidth]{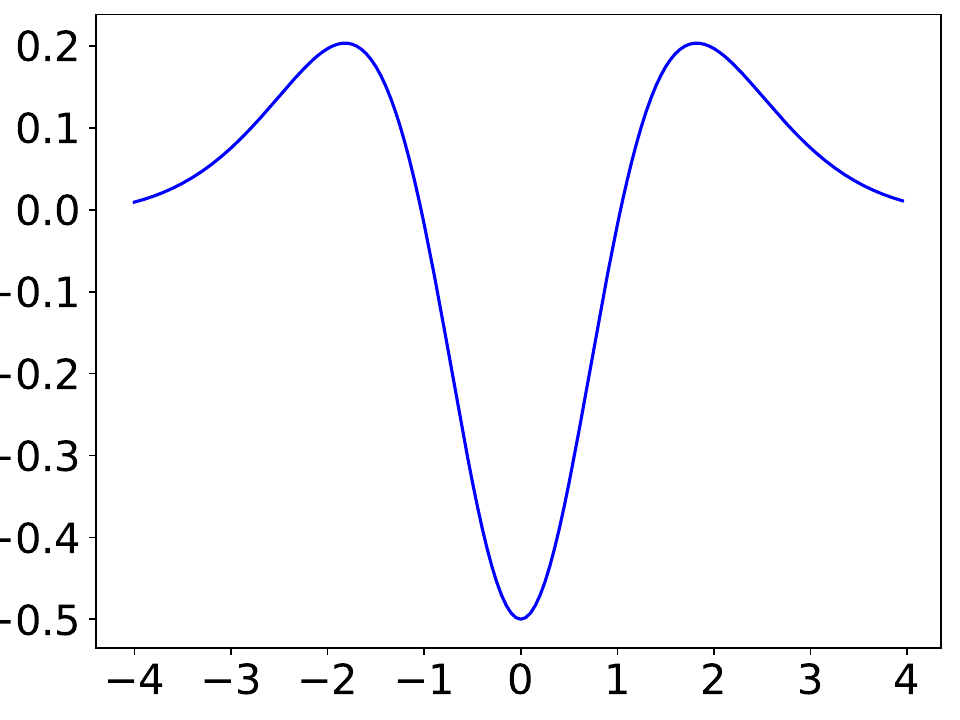}
	\end{minipage}
}
\subfigure[$y=\text{silu}^{(5)}(x)$]{
\centering
	\begin{minipage}[t]{0.3\linewidth}
		\includegraphics[width=0.99\linewidth]{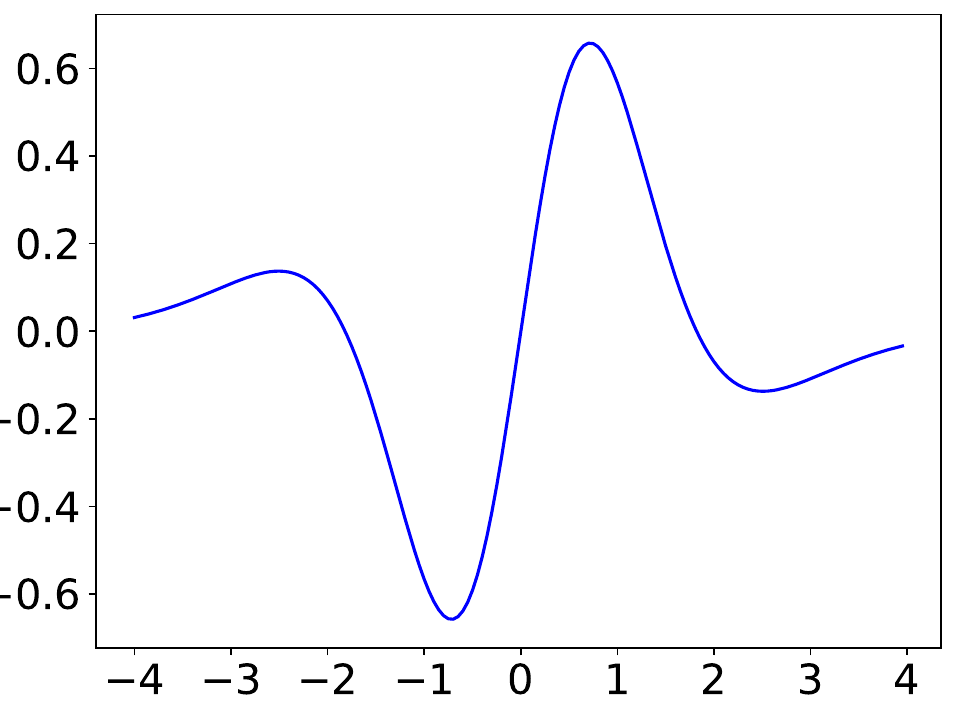}
	\end{minipage}%
}
\caption{Plot of $y=\text{silu}^{(5)}(x)$, where $n = 0,1,2,3,4,5$.}
\label{fig:silu_grad}
\end{figure*}

\noindent
According to the Leibniz rule, we have the $n$-order derivative of $\text{silu}(\bullet)$ given by
\begin{align}
    \text{silu}^{(n)}(x) &= \sum_{k=0}^n \binom{k}{n} x^{(k)} \sigma^{(n-k)} (x)
    \nonumber
    \\
    &= x \sigma^{(n)}(x) + n \sigma^{(n-1)}(x).
\end{align}
We let
\begin{align}
    &h_n (x) = \sigma^{(n)}(x)
    \nonumber
    \\
    &= [\sigma(x) (1 - \sigma(x))]^{(n-1)}
    \nonumber
    \\
    &= \sum_{k=0}^{n-1} \binom{k}{n-1} \sigma^{(k)}(x) (1 - \sigma(x))^{(n-k)}
    \nonumber
    \\
    &= \sum_{k=0}^{n-1} \binom{k}{n-1} \sigma^{(k)}(x) (- \sigma(x))^{(n-k-1)}
    \nonumber
    \\
    \label{eq:silu_h_n}
    &= -\sum_{k=0}^{n-1} \binom{k}{n-1} h_{k}(x) h_{n-k-1}(x).
\end{align}
The $n$-order derivative of $\text{silu}(\bullet)$ is given by
\begin{equation}
    \text{silu}^{(n)}(x) = x h_{n}(x) + n h_{n-1}(x),
\end{equation}
where $h_n (x)$ can be recursively computed by Equation~(\ref{eq:silu_h_n}); $h_0(x) = \sigma(x)$, $h_1(x) = \sigma(x)(1-\sigma(x))$.
We also visualize $y=\text{silu}^{(n)}(x)$ with $n = 0,1,2,3,4,5$ in Figure~\ref{fig:silu_grad}~(a)-(f), respectively.


\section{Details about Datasets}

All open-sourced datasets have the Apache-2.0 licence, which allows for academic research.

\paragraph{TruthfulQA:} TruthfulQA is a benchmark designed to assess the truthfulness of a language model's answers. The dataset consists of 4,114 questions across 38 categories, including health, law, finance, and politics. These questions are crafted to reflect common false beliefs or misconceptions that humans might hold~\cite{lin2021truthfulqa}.

\paragraph{MathQA:} MathQA is a comprehensive dataset of math word problems accompanied by an interpretable neural solver that translates problems into operational programs. It contains 14,925 questions~\cite{amini-etal-2019-mathqa}.

\paragraph{MMLU:} The MMLU benchmark measures the knowledge acquired during pretraining by evaluating models in zero-shot and few-shot settings. It includes 56,168 questions covering 57 subjects across STEM, humanities, social sciences, and more, with difficulty levels ranging from elementary to advanced professional. The benchmark tests both general knowledge and problem-solving abilities~\cite{hendryckstest2021}.

\paragraph{OpenbookQA:} OpenbookQA addresses the task of open-domain question answering using datasets like Wikipedia. This dataset contains 2,000 questions~\cite{OpenBookQA2018}.

\paragraph{C4:} C4 is a large, cleaned version of the Common Crawl web corpus. For LLM distillation in Section~\ref{sec:exp_distill}, we use the 'c4-train.00000-of-01024.json.gz' split, and down-sample a subset of 35.6k instances for the distillation.

\paragraph{Wikitext-2:} The WikiText-2 dataset is a collection of over 100 million tokens from verified Good and Featured articles on Wikipedia. It includes 4.4k sentences used to evaluate the perplexity of language models.


\section{Implementation Details}
\label{appendix:implement_details}

When selecting the protected columns within the $\texttt{down\_projection}$ weights, we aim to minimize the difference between the embedding $\textbf{z}$ and the local embedding $\textbf{z}_0$, represented by $\textbf{z} - \textbf{z}^*_0$. Specifically, according to Equation~(\ref{eq:local_embedding}), the approximate upper bound for each dimension of $\textbf{z} - \textbf{z}^*_0$ is given by $|\textbf{z}_{\text{max}} - \textbf{z}_{\text{min}}|$, where $|\bullet|$ takes the element-wise absolute values. 
Then, the indexes of the protected columns take the least $K$ dimensions within $|\textbf{z}_{\text{max}} - \textbf{z}_{\text{min}}|$, where $K$ takes $1 \times 10^4$, $8 \times 10^3$, and 2560 for the Llama-3-8B, Mistral-7B, and Phi-2 LLMs, respectively.
The number of protected columns takes the maximal value that \Algnameabbr{} can retain the original performance of the LLMs.
Other Experimental configurations are in Table~\ref{tab:computing_infrastructure}.
The unprotected columns preserve their pre-trained values in the fine-tuning (Sections~\ref{sec:exp_ft}) and distillation experiments (~\ref{sec:exp_distill}).

\section{Computational Infrastructure}
\label{appendix:hardware}

The computational infrastructure information is given in Table~\ref{tab:computing_infrastructure}.

\begin{table}[H]
\centering
\caption{Experiment configuration and computing infrastructure.}
\begin{tabular}{l|c}
\toprule
Name & Value \\
\midrule
Data type & \texttt{torch.bfloat16} \\
Flash-Attention & True \\
Eval batch-size & 1 \\
Computing Infrastructure & GPU \\
GPU Model & NVIDIA-A100 \\ 
GPU Memory & 80GB \\ 
GPU Number & 8 \\
CUDA Version & 12.1 \\
CPU Memory & 512GB \\
\bottomrule
\end{tabular}
\label{tab:computing_infrastructure}
\end{table}

\section{Hyper-parameters for Fine-tuning and Distillation}
\label{appendix:ft_hyper}

Table~\ref{tab:ft_hyper} shows the hyper-parameter settings of fine-tuning and distillation experiments in Sections~\ref{sec:exp_ft} and~\ref{sec:exp_distill}.

\begin{table}[H]
\centering
\caption{Hyper-parameter settings of fine-tuning and distillation experiments in Sections~\ref{sec:exp_ft} and~\ref{sec:exp_distill}.}
\begin{tabular}{l|c}
\toprule
Name & Value \\
\midrule
Optimizer & AdamW \\
Learning rate & $10^{-5}$ \\
LR scheduler & Linear \\
Warmup steps & 0 \\
Mini-batch size & $8 \times 8$ \\
Fine-tuning Epoch & 10 \\
Distillation Epoch & 1 \\
\bottomrule
\end{tabular}
\label{tab:ft_hyper}
\end{table}

\section{Packages}

In this work, we use the \texttt{transformers} along with \texttt{datasets} packages~\cite{wolf2019huggingface} for model and dataset loading, and \texttt{lm-eval-harness}~\cite{gao10256836framework} package for evaluation.
All open-sourced packages have the Apache-2.0 licence, which allows for academic research.

\end{document}